\documentclass[sigconf,screen]{acmart}
\usepackage{microtype}
\usepackage[utf8]{inputenc}
\usepackage{xspace}
\usepackage{graphicx}
\usepackage{listings}
\usepackage{xcolor}
\definecolor{darkgreen}{rgb}{0.0, 0.5, 0.0}
\definecolor{darkred}{rgb}{0.7, 0.0, 0.0}
\usepackage{booktabs}
\usepackage{multirow}
\usepackage{subcaption}
\usepackage{tcolorbox}
\usepackage{stfloats}
\usepackage{placeins}
\usepackage{float}
\tcbuselibrary{skins}
\newtcolorbox[auto counter]{findingbox}{
  enhanced,
  colback=gray!10,
  colframe=gray!50,
  fonttitle=\bfseries,
  title=Finding~\thetcbcounter,
  left=6pt, right=6pt, top=4pt, bottom=4pt,
}

\AtBeginDocument{%
  }

\newcommand{\approach}{{SeeRepo}\xspace}
\newcommand{\rqonefull}{How effective are current MLLMs at repository-level issue resolution?}
\newcommand{\rqtwofull}{How to integrate MLLMs into current agentic frameworks for repository-level issue resolution?}
\newcommand{\rqthreefull}{How do different visual layouts affect MLLMs' ability for repository issue resolution?}
\newcommand{\rqfourfull}{Which stage are visual tools most effective when utilized for software issue resolution?}

\setcopyright{cc}
\setcctype{by}
\acmDOI{10.1145/3832783.3834363}
\acmYear{2026}
\copyrightyear{2026}
\acmISBN{979-8-4007-2882-2/2026/10}
\acmConference[ASE '26]{Proceedings of the 41st IEEE/ACM International Conference on Automated Software Engineering}{October 12--16, 2026}{Munich, Germany}
\acmBooktitle{Proceedings of the 41st IEEE/ACM International Conference on Automated Software Engineering (ASE '26), October 12--16, 2026, Munich, Germany}
\acmSubmissionID{ase26main-p734-p}
\received{2026-03-26}
\received[accepted]{2026-06-18}

\begin{document}

\title{LLM Agents Can See Code Repositories}
\author{Dongjian Ma}
\orcid{0009-0006-7996-492X}
\authornote{Dongjian Ma and Silin Chen contributed equally to this work.}
\affiliation{%
  \institution{Shanghai Jiao Tong University}
  \city{Shanghai}
  \country{China}
}
\email{mdj200226@gmail.com}

\author{Silin Chen}
\authornotemark[1]
\orcid{0009-0001-9914-4172}
\affiliation{%
  \institution{Shanghai Jiao Tong University}
  \city{Shanghai}
  \country{China}
}
\email{csl2457029646@163.com}

\author{Yufei Yang}
\orcid{0009-0001-1621-0497}
\affiliation{%
  \institution{Xi'an Jiaotong University}
  \city{Xi'an}
  \country{China}
}
\email{qfrfyflc@stu.xjtu.edu.cn}

\author{Yuling Shi}
\orcid{0009-0009-9738-7072}
\affiliation{%
  \institution{Shanghai Jiao Tong University}
  \city{Shanghai}
  \country{China}
}
\email{yuling.shi@sjtu.edu.cn}

\author{Yanfu Yan}
\orcid{0009-0008-2475-6802}
\affiliation{%
  \institution{Zhejiang University,}
  \city{}
  \country{}
}
\affiliation{%
  \institution{Hangzhou High-Tech Zone (Binjiang) Institute of Blockchain and Data Security}
  \city{Hangzhou}
  \country{China}
}
\email{yanfu@zju.edu.cn}

\author{Xiaodong Gu}
\authornote{Corresponding author.}
\orcid{0000-0002-0529-6408}
\affiliation{%
  \institution{Shanghai Jiao Tong University}
  \city{Shanghai}
  \country{China}
}
\email{xiaodong.gu@sjtu.edu.cn}

\renewcommand{\shortauthors}{Dongjian Ma, Silin Chen, Yufei Yang, Yuling Shi, Yanfu Yan, and Xiaodong Gu}

\begin{abstract}
Coding agents powered by large language models (LLMs) have demonstrated remarkable proficiency in software engineering tasks. Yet modern coding agents rely almost entirely on text, leaving a major gap between how human developers and agents comprehend software repositories. Human developers leverage visual repository cues such as folder hierarchies and dependencies, raising the question of whether multimodal models can similarly improve repository understanding.

In this paper, we conduct the first systematic empirical study on multimodal foundation models for repository-level tasks. Our experiments across four modern multimodal models reveal that while a vision-only context representation degrades performance and inflates token costs, integrating visualized context graphs as a supplementary modality can help agents grasp the repository more efficiently. Specifically, providing agents with visual structural context alongside standard text interfaces reduces input token consumption by up to 26\% while maintaining or improving issue-resolution accuracy. Furthermore, we demonstrate that visual tools are most effective when utilized during the fault localization stage and when agents autonomously dictate their exploration depth. Our findings highlight a promising hybrid-modality pathway for the design of next-generation coding agents.
\end{abstract}



\maketitle

\begin{figure}[t]
  \centering
  \includegraphics[width=\linewidth]{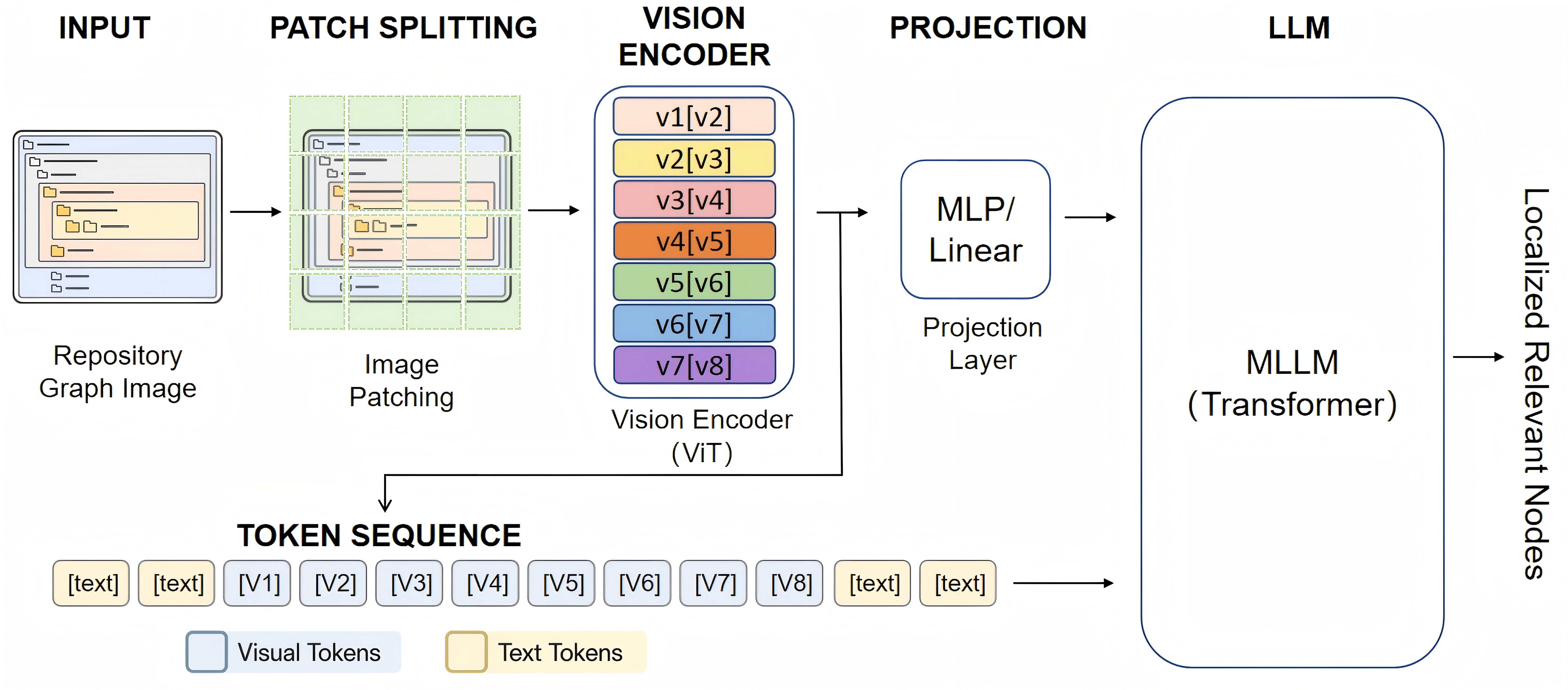}
  \caption{Process of how MLLMs perceive multimodal rendering of a code repository.}
  \label{fig:background}
\end{figure}

\section{Introduction}

Coding agents empowered by large language models (LLMs) have shown remarkable capabilities in software engineering tasks, including resolving issues in large repositories~\cite{yang2024sweagenta,jimenez2023swe, zeng2025pruning, zeng2026dockerless, zeng2026glimprouter, hu2026line, gao2026swe}. Most existing agents interact with repositories through tokenized text: source code, documentation, and execution feedback are flattened into sequences for reasoning and planning. While this text-centric paradigm has driven rapid progress, it raises an open question: \emph{Is text the most effective modality for presenting repository context to modern foundation models?}

To developers, code repositories comprise artifacts from multiple modalities, including not only textual source code and documentation but also structural relationships such as function dependencies, \textit{etc.}. However, existing techniques for repository understanding across a range of SE tasks typically rely on linearizing these heterogeneous artifacts into sequential inputs. As a result, models are required to infer structural information that was originally conveyed through non-linear or visual representations. Recovering such organizations and relationships can be challenging under limited context budgets. While prior work has explored structural representations of code repositories, such as graph-based abstractions~\cite{liu2024codexgraph, chen2025locagent, jiang2025cosil}, the information ultimately consumed by models at inference time remains predominantly in the form of text tokens. Even when graph encodings are used, they are typically linearized for model input~\cite{liu2024graphcoder}, which may lead to the loss of important relational cues. In contrast, visual representations of repositories can expose additional signals—such as two-dimensional layout and stable spatial grouping—that are not naturally captured by linear text. More broadly, visual context may provide richer information per unit of prompt, potentially improving an agent’s ability to remain oriented, retrieve relevant context, and perform accurate edits in long-horizon repository workflows.

In this paper, we conduct the first empirical study on multimodal foundation models for repository-level tasks. We study the effects of shifting repository representations toward visual modalities. Specifically, we evaluate four multimodal models—GPT-5-mini~\cite{openai2025gpt5}, GPT-5.1~\cite{openai2025gpt51}, Doubao-Seed-2.0-Lite~\cite{bytedance2026doubao}, and Kimi K2.5~\cite{moonshot2026kimi}—and analyze how design choices—including representation modality, image–text balance, visualization layout design, and visualization tool invocation stages—affect agent performance and efficiency. To support this study, we build \approach, a multimodal augmentation for coding agents on repository-level issue resolution. It integrates visual graph renderings of repository structure with standard text-based code access and editing, enabling agents to combine structural awareness from vision with symbolic precision from text. Concretely, \approach{} uses AST-based static analysis to construct multi-relation dependency graphs capturing containment, import, invocation, and inheritance relationships. Given a query node, it renders a Graphviz subgraph centered on that node as a PNG image, which the agent receives alongside conventional text-based code access. This hybrid interface allows the agent to leverage spatial structure from vision while retaining symbolic precision from text. Our study is guided by four research questions:

\begin{figure*}[t]
  \centering
  \includegraphics[width=0.95\textwidth, trim=10 30 10 0 clip]{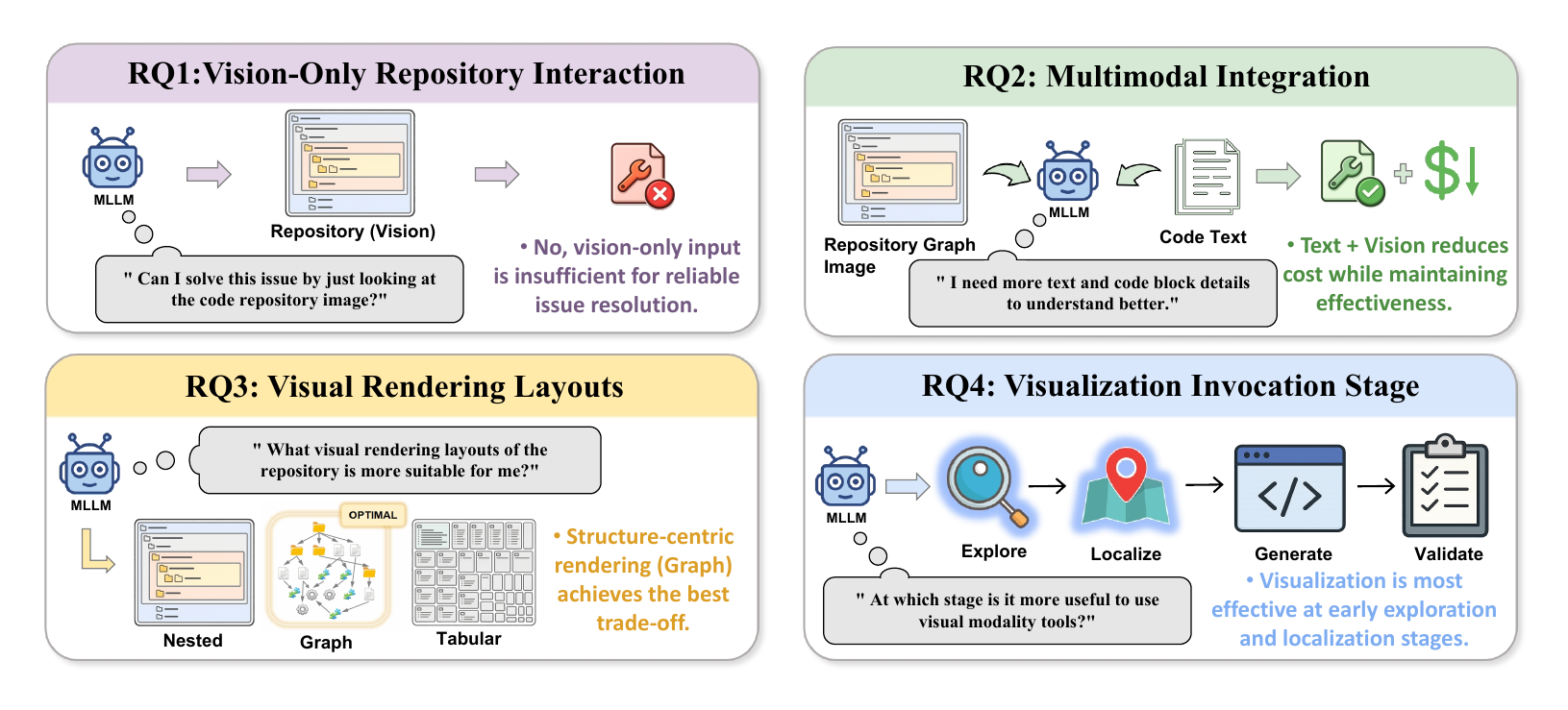}
  \caption{Overview of the Study Design and Core Findings}
  \label{fig:overview}
\end{figure*}

\noindent\textbf{RQ1: \rqonefull}
We investigate whether visual representations are effective in repository-level understanding. We set the Mini-SWE-Agent in a vision-only mode, where bash navigation commands return graph images instead of texts. We evaluate three representative MLLMs and compare the performance of vision-only modality against traditional text modality.

\textbf{Findings.}
Vision-only interaction significantly degrades resolution accuracy across all three models: GPT-5-mini drops from 55.0\% to 41.4\% ($-$13.6), Doubao-Seed-2.0-Lite drops sharply from 51.0\% to 16.9\% ($-$34.1), and Kimi K2.5 drops from 70.3\% to 55.0\% ($-$15.3). Contrary to expectations, token cost surges rather than decreases---GPT-5-mini incurs 42\% higher cost, Doubao exhibits 268\% cost inflation, and Kimi K2.5 sees a 27\% increase. Agents deprived of text access resort to repeated graph queries to compensate for missing symbolic information, accumulating high overhead without recovering accuracy. This suggests that current MLLMs are heavily reliant on symbolic cues for efficient code reasoning~\cite{wei2022chain}, and raw visual graphs alone cannot provide sufficient semantic guidance.

\noindent\textbf{RQ2: \rqtwofull}
Having observed that vision-only repository interaction degrades the performance of issue resolution, we wonder whether a hybrid \emph{text+vision} context can combine the strengths of both modalities and thus enhance the performance of coding agents. We integrate the vision representation of the code repository as a supplementary context alongside the standard bash-based interface, and compare the performance under various models.

\textbf{Findings.}
Integrating \approach{} as a supplementary modality significantly reduces the agent's token cost while maintaining or improving resolution accuracy. On GPT-5-mini, Pass@1 improves to 55.4\% (+0.4) with input tokens reduced by 25\% and cost by 26\%. GPT-5.1 achieves a 46\% cost reduction despite a minor accuracy dip ($-$2.2). Kimi~K2.5 simultaneously improves Pass@1 by 1.8 points (68.8\%$\to$70.6\%) and reduces cost by 3\%. Doubao-Seed-2.0-Lite gains +1.0 Pass@1 with a 6\% cost reduction. The same trend also transfers to additional GPT-5-mini evaluations on SWE-Rebench Leaderboard 2026.03~\cite{badertdinov2026swe} and SWE-QA~\cite{peng2025swe}. The results suggest that multi-modal repository representation can help agents grasp code context more efficiently.

\noindent\textbf{RQ3: \rqthreefull}
Having established that multimodal representation helps coding agent understand repository more efficiently, we examine which visual rendering strategy yields the best efficiency.
We experiment with three visual rendering strategies (graph, nested, and tabular) at different hierarchical depths (fixed vs. dynamic) and compare with the text-only baseline. 

\textbf{Findings.}
All three visual layouts outperform the text-only baseline in token efficiency. Graph-based layout achieves the greatest token reduction ($-$25\% input tokens, $-$26\% cost) with a Pass@1 of 55.4\% (+0.4); nested and tabular layouts trade slightly higher token cost for marginal accuracy gains of +0.8 and +1.2 respectively. In terms of hierarchical depths, the dynamic depth strategy achieves a competitive Pass@1 gain (+0.4) while delivering the largest reductions in input tokens ($-$25\%) and cost ($-$26\%) across all depth configurations.

\noindent\textbf{RQ4: \rqfourfull}
We integrate vision representations into one of the three stages in the issue resolution pipeline---localization, repair, and patch validation, respectively --- and analyze how visualization affects performance in each stage.

\textbf{Findings.}
Visualization is most effective when invoked at the fault localization stage: a multimodal agent equipped with \approach{} achieves Pass@1 of 55.4\% (+0.4), reduces input tokens by 25\%, and lowers cost by 26\%.

To sum up, this paper makes the following contributions:
\begin{itemize}
    \item We present the first systematic, large-scale study of visual repository representations for coding agents on the software issue resolution task, analyzing how modality, layout, and invocation stage affect effectiveness and efficiency across four modern multimodal models.

    \item We empirically demonstrate a performance boundary between vision-only and multimodal repository representations: while vision-only repository representation is ineffective for issue resolution, combining both textual and visual representations yields a substantially better trade-off between effectiveness and efficiency.

    \item We design and implement \approach as an experimental framework for studying repository visualization in coding agents. Through extensive experiments, we further show that structure-centric renderings and early-stage visualization during exploration and localization are the most effective ways to leverage visual context.
\end{itemize}

\section{Background}

\subsection{Multimodal Large Language Models}

Multimodal Large Language Models (MLLMs) extend traditional language models by incorporating visual perception, transforming images into sequences of visual tokens that can be jointly processed with text via a unified Transformer architecture~\cite{dosovitskiy2020image}. As illustrated in Figure~\ref{fig:background}, an input image $\mathbf{I} \in \mathbb{R}^{H \times W \times 3}$ is divided into $N$ fixed-size patches $\{p_i\}_{i=1}^{N}$, each encoded by a vision encoder (e.g., ViT) into a dense visual embedding $\mathbf{v}_i \in \mathbb{R}^{d_v}$. A projection layer $f_\theta$, typically learned, maps these embeddings into the language model's token space:
\begin{equation}
  \tilde{\mathbf{v}}_i = f_\theta(\mathbf{v}_i) \in \mathbb{R}^{d},
\end{equation}
producing a visual token sequence $[\tilde{\mathbf{v}}_1, \ldots, \tilde{\mathbf{v}}_N]$ that is concatenated with text tokens and processed by the Transformer. The 2D spatial arrangement of patches is approximately preserved via positional embeddings, enabling cross-modal attention to align visual regions with textual semantics. Consequently, MLLMs can leverage visual layouts to capture global context and relational structure that are difficult to represent purely through linear text sequences, even in software engineering contexts~\cite{cheng2024seeclick,ye2023mplug}.

\subsection{Visual Perception of Repository Structure}

Software repositories naturally exhibit rich topological structures, including dependency graphs, call relations, and modular hierarchies, which encode global context about program organization. Formally, a repository can be modeled as a directed heterogeneous graph $\mathcal{G} = (\mathcal{V}, \mathcal{E}, \mathcal{A}, \mathcal{R})$~\cite{chen2025locagent}, where nodes $v \in \mathcal{V}$ represent files, classes, and functions with type $\phi(v) \in \mathcal{A}$, and edges $(u, v, r) \in \mathcal{E}$ capture typed relationships $r \in \mathcal{R} = \{\textit{contains},\, \textit{imports},\, \textit{inherits},\, \allowbreak \textit{invokes}\}$. Multiple edge types are allowed between node pairs.

Prior approaches predominantly serialize such structural information into text~\cite{liu2024codexgraph,liu2024graphcoder,jiang2025cosil}, which can obscure higher-order relationships and introduce substantial token overhead. Recent work has begun exploring visual modalities for software engineering tasks~\cite{huang2025seeingisfixing,tang2026svrepair}, but a systematic study of visual repository representations remains absent. By rendering $\mathcal{G}$ as an image, spatial locality and connectivity patterns are preserved: node positions encode module membership, edge trajectories encode dependency direction, and cluster boundaries in the rendering typically correspond to architectural boundaries. Treating repository graphs as images allows MLLMs to leverage their visual reasoning capabilities to potentially perceive structural dependencies and global organization, enabling structure-aware understanding that complements textual code representations.

\begin{table*}[!t]
  \centering
  \caption{Effect of MLLMs on SWE-bench Verified (500 instances).}
  \label{tab:rq1_vision_only}
  \small
  \setlength{\tabcolsep}{4pt}
  \begin{tabular}{l c | cccc}
  \toprule
  \multirow{2}{*}{Agent} & \textbf{Effectiveness} & \multicolumn{4}{c}{\textbf{Efficiency}} \\
  \cmidrule{2-6}
  & Pass@1 $\uparrow$ & API Calls $\downarrow$ & Input Tokens $\downarrow$ & Output Tokens $\downarrow$ & Cost (\$) $\downarrow$ \\
  \midrule
  \multicolumn{6}{c}{\textbf{GPT-5-mini}} \\
  \midrule
  Text
      & 55.0\% & 15 & 193,157 & 8,188 & 0.031 \\
  Vision-Only
      & 41.4\% \textcolor{darkred}{(-13.6)} & 20 \textcolor{darkred}{(+33\%)} & 270,117 \textcolor{darkred}{(+40\%)} & 12,352 \textcolor{darkred}{(+51\%)} & 0.044 \textcolor{darkred}{(+42\%)} \\
  \midrule
  \multicolumn{6}{c}{\textbf{Doubao-Seed-2.0-Lite}} \\
  \midrule
  Text
      & 51.0\% & 22 & 201,754 & 5,562 & 0.019 \\
  Vision-Only
      & 16.9\% \textcolor{darkred}{(-34.1)} & 28 \textcolor{darkred}{(+27\%)} & 965,272 \textcolor{darkred}{(+379\%)} & 6,214 \textcolor{darkred}{(+12\%)} & 0.070 \textcolor{darkred}{(+268\%)} \\
  \midrule
  \multicolumn{6}{c}{\textbf{Kimi K2.5}} \\
  \midrule
  Text
      & 70.3\% & 40 & 639,474 & 9,775 & 0.1213 \\
  Vision-Only
      & 55.0\% \textcolor{darkred}{(-15.3)} & 78 \textcolor{darkred}{(+95\%)} & 879,853 \textcolor{darkred}{(+38\%)} & 9,258 \textcolor{darkgreen}{(-5\%)} & 0.1543 \textcolor{darkred}{(+27\%)} \\
  \bottomrule
  \end{tabular}
\end{table*}

\section{Experimental Setup}
\subsection{Benchmark and Metrics}
We conduct our main experiments on \textbf{SWE-bench Verified}~\cite{openai2024swebenchverified}, a human-curated subset of SWE-bench~\cite{jimenez2023swe} released by OpenAI to provide more reliable and reproducible evaluations of AI agents on real-world software engineering tasks. Unlike the original benchmark, each instance in SWE-bench Verified has been manually inspected to ensure task validity, making it a widely used standard for assessing autonomous coding agents. The benchmark comprises 500 instances drawn from widely-used Python projects on GitHub, where each instance presents a real bug report paired with a set of unit tests that determine whether a submitted patch correctly resolves the underlying issue.
We additionally evaluate GPT-5-mini transfer performance on two external benchmarks: SWE-Rebench Leaderboard~\cite{badertdinov2026swe} (110 instances across 41 repositories from the 2026.03 release) and SWE-QA~\cite{peng2025swe}, a repository-level code question answering benchmark. Together, they complement SWE-bench Verified by testing whether visual structural grounding generalizes to both new issue-resolution tasks and evidence-heavy repository QA scenarios.

To evaluate the performance and efficiency of the multimodal models under different design choices, we select the following metrics:

\begin{itemize}
  \item \textbf{Pass@1:} The percentage of issue-resolution benchmark instances for which the multimodal model generates a patch that successfully resolves the task. We report Pass@1 for SWE-bench Verified and SWE-Rebench.
  \item \textbf{Overall Score:} The official aggregate score used by SWE-QA, reported on a 0--100 scale. This metric captures answer quality on repository-level code question answering tasks.
  \item \textbf{Number of API Calls:} The average number of API calls made per benchmark instance, reflecting the number of interaction steps the agent takes to resolve an issue.
  \item \textbf{Number of Input Tokens:} The average number of input tokens consumed per benchmark instance, capturing the context overhead introduced by the agent's prompts and visual inputs.
  \item \textbf{Number of Output Tokens:} The average number of completion tokens generated per benchmark instance, reflecting the verbosity of the agent's responses.
  \item \textbf{Average Cost per Instance:} The average monetary cost incurred to process a single benchmark instance, computed by dividing the total evaluation cost by the number of evaluated instances.
\end{itemize}


\subsection{Implementation Details}
Our framework is built on top of Mini-SWE-Agent~\footnote{\url{https://github.com/swe-agent/mini-swe-agent}} and extends its tool-calling interface with a repository graph visualization module. For each target repository, we pre-construct a directed heterogeneous graph with four relation types—\textit{contains}, \textit{imports}, \textit{invokes}, and \textit{inherits}—and serialize it for reuse during inference. During inference, the agent queries the graph via an external tool, which renders the requested subgraph as a PNG image using Graphviz and returns it as visual context. All models are queried with temperature 0. Each agent run is limited to 250 interaction steps and a cost budget of \$3.0 per instance. Unless otherwise specified, all reported results are averaged over three independent runs under this setting.

\begin{table*}[!t]
  \centering
  \caption{Performance comparison between text and multimodal agents (\approach{}) across models.}
  \small
  \setlength{\tabcolsep}{4pt}
  \begin{tabular}{l c | cccc}
  \toprule
  \multirow{2}{*}{Agent} & \textbf{Effectiveness} & \multicolumn{4}{c}{\textbf{Efficiency}} \\
  \cmidrule{2-6}
  & Pass@1 $\uparrow$ & API Calls $\downarrow$ & Input Tokens $\downarrow$ & Output Tokens $\downarrow$ & Cost (\$) $\downarrow$ \\
  \midrule
  \multicolumn{6}{c}{\textbf{GPT-5-mini}} \\
  \midrule
  Text
      & 55.0\% & 15 & 193,157 & 8,188 & 0.031 \\
  Multimodal
      & \textbf{55.4\%} \textcolor{darkgreen}{(+0.4)} & \textbf{13} \textcolor{darkgreen}{(-13\%)} & \textbf{144,403} \textcolor{darkgreen}{(-25\%)} & \textbf{6,958} \textcolor{darkgreen}{(-15\%)} & \textbf{0.023} \textcolor{darkgreen}{(-26\%)} \\
  \midrule
  \multicolumn{6}{c}{\textbf{GPT-5.1}} \\
  \midrule
  Text
      & \textbf{51.0\%} & 16 & 206,256 & 3,113 & 0.1795 \\
  Multimodal
      & 48.8\% \textcolor{darkred}{(-2.2)} & \textbf{13} \textcolor{darkgreen}{(-19\%)} & \textbf{161,130} \textcolor{darkgreen}{(-22\%)} & \textbf{2,342} \textcolor{darkgreen}{(-25\%)} & \textbf{0.0975} \textcolor{darkgreen}{(-46\%)} \\
  \midrule
  \multicolumn{6}{c}{\textbf{Kimi K2.5}} \\
  \midrule
  Text
      & 68.8\% & 41 & \textbf{691,141} & 10,210 & 0.1270 \\
  Multimodal
      & \textbf{70.6\%} \textcolor{darkgreen}{(+1.8)} & \textbf{40} \textcolor{darkgreen}{(-2\%)} & 723,874 \textcolor{darkred}{(+5\%)} & \textbf{9,265} \textcolor{darkgreen}{(-9\%)} & \textbf{0.1229} \textcolor{darkgreen}{(-3\%)} \\
  \midrule
  \multicolumn{6}{c}{\textbf{Doubao-Seed-2.0-Lite}} \\
  \midrule
  Text
      & 51.0\% & 22 & 185,311 & 5,383 & 0.0173 \\
  Multimodal
      & \textbf{52.0\%} \textcolor{darkgreen}{(+1.0)} & \textbf{21} \textcolor{darkgreen}{(-4\%)} & \textbf{176,463} \textcolor{darkgreen}{(-4.8\%)} & \textbf{4,958} \textcolor{darkgreen}{(-7.9\%)} & \textbf{0.0162} \textcolor{darkgreen}{(-6.0\%)} \\
  \bottomrule
  \end{tabular}
  \label{tab:mini_swe_agent_approach}
\end{table*}

\section{RQ1: Effectiveness of Current MLLMs at Issue Resolution Tasks}

We first investigate whether MLLMs can resolve repository-level issues. Specifically, we capture the outputs of the agent's Bash navigation and file-reading commands to images and feed them to the MLLM. 

\subsection{Experimental Design}
In the vision-only setting, the agent operates with its standard tool interface, but all Bash commands (e.g., \texttt{cat}, \texttt{find}, \texttt{grep}) return visual graph images rendered by \approach{} instead of text output. The agent must navigate the repository, identify relevant code entities, and generate patches by interpreting these visual responses. We evaluate this setting on GPT-5-mini, Doubao-Seed-2.0-Lite, and Kimi K2.5 using the full SWE-bench Verified.

\subsection{Result Analysis}
As shown in Table~\ref{tab:rq1_vision_only}, vision-only context representation substantially reduces the accuracy for all three models. GPT-5-mini drops from 55.0\% to 41.4\% ($-$13.6), Doubao-Seed-2.0-Lite drops sharply from 51.0\% to 16.9\% ($-$34.1), and Kimi K2.5 drops from 70.3\% to 55.0\% ($-$15.3). All models also incur significantly higher costs: GPT-5-mini sees a 42\% cost increase, Doubao 268\%, and Kimi K2.5 27\%. Notably, Kimi K2.5 nearly doubles its API calls (+95\%), suggesting it compensates for the lack of textual information through more frequent but shorter queries.

When deprived of text representations, agents resort to repeated graph queries to compensate for missing precise symbolic information, accumulating high token overhead without improving accuracy. The three models exhibit distinct strategies under this constraint: Doubao-Seed-2.0-Lite engages most extensively with visual graphs, leading to a 379\% input token surge and the steepest accuracy drop ($-$34.1); Kimi K2.5 adopts a high-frequency query strategy with 95\% more API calls but only 38\% more input tokens, maintaining relatively higher accuracy; while GPT-5-mini appears to abandon visual exploration earlier and generates patches with incomplete context. These results indicate that while MLLMs can process visual repository structure to varying degrees, graph images alone provide insufficient symbolic information for accurate issue resolution.

\begin{findingbox}
Vision-only repository interaction degrades issue resolution performance across all evaluated models: the resolution accuracy drops by 13.6 to 34.1 points, while token cost increases substantially (up to 268\% higher cost). Agents resort to repeated graph queries to compensate for the absence of textual information, with API calls increasing by up to 95\%.
\end{findingbox}

\section{RQ2: Effect of Multimodal Context Integration}

Having seen that vision-only repository representation is ineffective, we investigate whether visual repository representation as a supplementary modality can improve issue resolution when integrated with standard text modality. 

\subsection{Methodology}
To enable multimodal repository perception, we design \approach{}, a tool that augments coding agents with visual renderings of repository structure. \approach{} visually perceive the repository's dependency graph alongside its standard text-based interface. 

Initially, \approach constructs a dependency graph for the repository via AST-based static analysis and renders query-centered subgraphs.
Following LocAgent~\cite{chen2025locagent}, we consider four types of directed relationships in the repository graph: \textit{contains} (filesystem hierarchy), \textit{imports} (module-level dependencies), \textit{inherits} (class inheritance chains), and \textit{invokes} (function-level call relationships). The graph is queriable at runtime by node identifier, edge type, and traversal depth in both upstream and downstream directions.

Next, \approach transforms the dependency graph into structured visual representations. When the agent decides to query a node from the repository, it first constructs a bidirectional graph. This graph is built by performing breadth-first traversal over upstream and downstream relations up to a specified depth. The process constructs a distance-aware subgraph centered on the query target, where upstream dependencies are assigned negative distances and downstream dependencies are assigned positive distances. This arrangement captures both dependency flow and structural proximity, positioning each node according to its relative distance from the query target.

The constructed subgraph is then rendered using a layered, left-to-right hierarchical layout generated by the Graphviz DOT engine\footnote{\url{https://graphviz.org/}}
. The rendered PNG images are capped at 35 million pixels to prevent excessive memory consumption for large subgraphs. Each node is displayed using a compact HTML-table label, augmented with semantic icons indicating entity types (e.g., files, modules, classes, or functions), while the queried node is visually highlighted to stabilize attention during reasoning. To reduce visual clutter in dense dependency regions, junction nodes are introduced to merge multiple outgoing edges before branching, improving edge readability without altering graph semantics. This layered visualization not only makes dependency flow explicit but also reveals structural proximity by placing closely related entities at similar distances from the query target.

\subsection{Experimental Setup}
We evaluate \approach on the bug localization task. The agent follows a structured three-phase localization strategy: (1) file hunting via the \textit{imports} graph, (2) logic hunting via the \textit{invokes} graph, and (3) hierarchy and path verification via \textit{inherits} and \textit{contains} as needed. Following the common localization pipeline, the agent reads relevant code snippets, implements a fix using standard Bash commands, and executes test cases to verify correctness. We evaluate \approach{} on 500 instances from SWE-bench Verified using four models: GPT-5-mini, GPT-5.1, Kimi K2.5, and Doubao-Seed-2.0-Lite. To assess whether the same trend transfers beyond SWE-bench Verified, we additionally evaluate GPT-5-mini on SWE-Rebench Leaderboard and SWE-QA. For SWE-Rebench Leaderboard, we use all 110 instances from the 2026.03 release, spanning 41 repositories. For SWE-QA, we follow the official evaluation protocol and report the Overall Score (0--100) together with the same efficiency statistics used in our main experiments.

\subsection{Results and Analysis}

As shown in Table~\ref{tab:mini_swe_agent_approach}, integrating multimodal context substantially reduces token consumption while improving or preserving accuracy: GPT-5-mini reduces total cost by 26\%, GPT-5.1 by 46\%, and Doubao-Seed-2.0-Lite by 6\%. This consistent reduction suggests that structural context primarily improves the \emph{localization phase} of issue resolution. In text-only interaction, agents typically rely on iterative file exploration, repeatedly issuing navigation commands to refine hypotheses about relevant code regions. By contrast, explicit structural grounding enables agents to narrow the candidate search space earlier, reducing redundant exploration and shortening reasoning trajectories. Concretely, a single graph query exposes the full dependency neighborhood of a target node in one step, short-circuiting the iterative grep-then-read cycle that otherwise requires multiple sequential navigation commands to accumulate equivalent structural context.
\begin{figure}[t]
  \centering
  \includegraphics[width=0.48\textwidth]{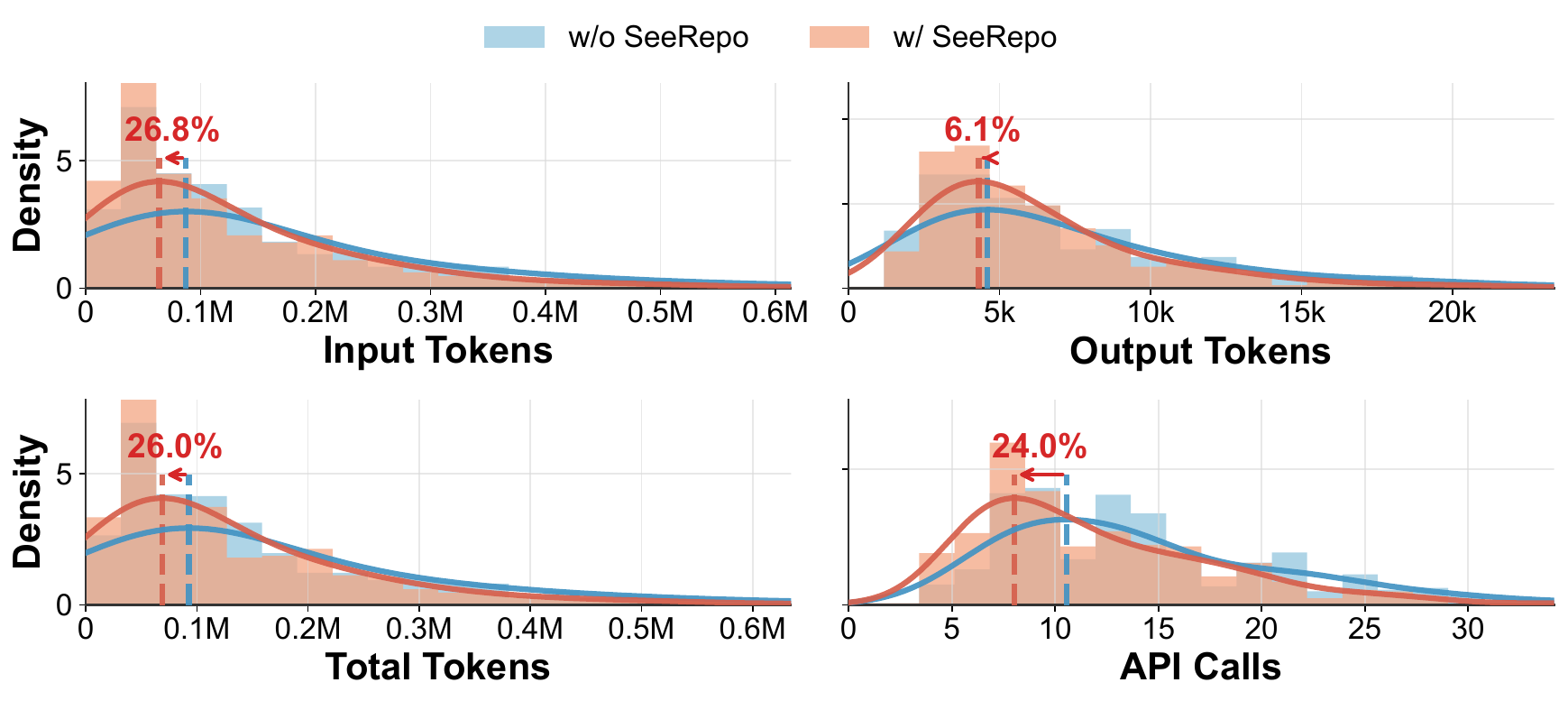}
  \caption{Efficiency analysis on SWE-bench Verified. {\color{orange}w/ \approach{}} achieves substantial reductions in both token cost and agent rounds compared to {\color{cyan}w/o \approach{}} for GPT-5-mini.}
  \label{fig:efficiency_analysis}
\end{figure}

The magnitude of efficiency gains varies across models, revealing differences in how models internalize structural signals. GPT-5.1 exhibits the largest cost reduction ($-$46\%) despite a slight accuracy drop ($-$2.2), suggesting that stronger base reasoning models already possess effective repair capabilities but benefit from structural context as a navigation prior that accelerates repository understanding. The accuracy regression likely reflects a tension specific to high-capability models: GPT-5.1's stronger parametric reasoning already allows it to form reliable localization hypotheses from sparse textual cues, so the additional graph queries occasionally redirect attention toward dependencies that are topologically proximate but semantically tangential to the defect. In this case, structural grounding trades a small amount of repair precision for substantially improved exploration efficiency.

Kimi K2.5 presents a contrasting pattern: input tokens increase slightly (+5\%), yet accuracy improves and overall cost still decreases by 3\%. Rather than using graph queries to replace textual exploration, Kimi K2.5 appears to treat them as supplementary context that reinforces its ongoing hypothesis formation: the near-unchanged API call count (41→40) alongside a 5\% input token increase suggests that structural and symbolic signals are processed in tandem rather than as substitutes. This more deliberative integration results in modestly longer inputs but more reliable downstream repair decisions. The outcome—highest Pass@1 (70.6\%) among all configurations—suggests that broadening context acquisition under coherent structural guidance can improve repair robustness even at the cost of some exploration overhead.

Doubao-Seed-2.0-Lite shows moderate accuracy gains with smaller efficiency improvements, suggesting that multimodal context primarily streamlines repository navigation without substantially changing the overall interaction pattern. Compared with other models, the reduction in token usage is more limited, indicating that visual representations mainly help agents reduce redundant exploration steps rather than reshaping the repair process.

\begin{figure*}[t]
  \centering
  \includegraphics[width=0.85\textwidth, trim=0 0 0 0 clip]{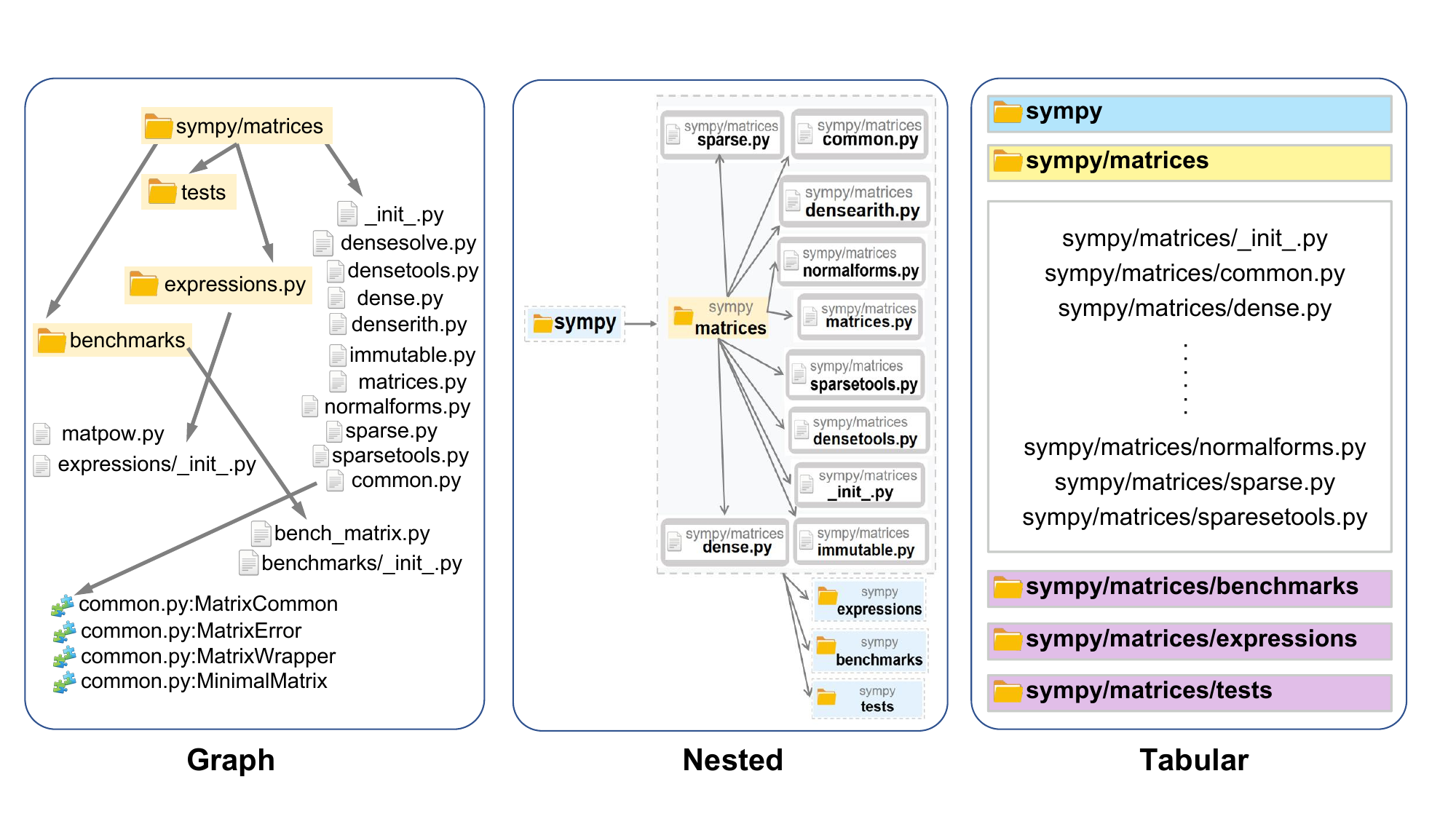}
  \caption{Examples of Three Visualization Styles.}
  \label{fig:layouts}
\end{figure*}

\begin{table}[t]
  \centering
  \caption{Additional GPT-5-mini evaluation on SWE-Rebench Leaderboard. The effectiveness metric is Pass@1 over the 110-instance 2026.03 release.}
  \scriptsize
  \setlength{\tabcolsep}{3pt}
  \resizebox{\columnwidth}{!}{%
  \begin{tabular}{l c | cccc}
    \toprule
    \multirow{2}{*}{Agent} & \textbf{Effectiveness} & \multicolumn{4}{c}{\textbf{Efficiency}} \\
    \cmidrule{2-6}
    & Pass@1 & API Calls & Input Tokens & Output Tokens & Cost (\$) \\
    \midrule
    Text & 25.45\% & 23 & 388,834 & 12,094 & 0.052 \\
    Multimodal & 26.36\%~{\textcolor{darkgreen}{(+0.91)}} & 21~{\textcolor{darkgreen}{(-2)}} & 253,132~{\textcolor{darkgreen}{(-34.89\%)}} & 10,242~{\textcolor{darkgreen}{(-15.32\%)}} & 0.047~{\textcolor{darkgreen}{(-9.6\%)}} \\
    \bottomrule
  \end{tabular}%
  }
  \label{tab:gpt5mini_swerebench}
\end{table}

\begin{table}[t]
  \centering
  \caption{Additional GPT-5-mini evaluation on SWE-QA. The effectiveness metric is the official Overall Score (0--100).}
  \scriptsize
  \setlength{\tabcolsep}{3pt}
  \resizebox{\columnwidth}{!}{%
  \begin{tabular}{l c | cccc}
    \toprule
    \multirow{2}{*}{Agent} & \textbf{Effectiveness} & \multicolumn{4}{c}{\textbf{Efficiency}} \\
    \cmidrule{2-6}
    & Score & API Calls & Input Tokens & Output Tokens & Cost (\$) \\
    \midrule
    Text & 66.8 & 14 & 66,369 & 5,003 & 0.0107 \\
    Multimodal & 67.2~{\textcolor{darkgreen}{(+0.4)}} & 9~{\textcolor{darkgreen}{(-35.7\%)}} & 50,038~{\textcolor{darkgreen}{(-24.6\%)}} & 4,254~{\textcolor{darkgreen}{(-15.0\%)}} & 0.0079~{\textcolor{darkgreen}{(-26.2\%)}} \\
    \bottomrule
  \end{tabular}%
  }
  \label{tab:gpt5mini_sweqa}
\end{table}

Figure~\ref{fig:efficiency_analysis} further illustrates efficiency differences on GPT-5-mini by comparing the text-only baseline with the multimodal context augmented with \approach{}. The multimodal setting exhibits a clear leftward shift in the distribution of input tokens, indicating that the visual representation of repository structure reduces the amount of repository context the agent needs to inspect before identifying relevant files.
Reductions in output tokens are smaller but consistent, suggesting that \approach{} primarily improves upstream exploration efficiency rather than substantially altering the verbosity of patch generation once a repair direction is established. This trend is reflected in total token usage, where the overall distribution shifts left by 26.0\%, confirming that efficiency gains are driven mainly by reduced context acquisition rather than shorter responses alone. Additionally, the multimodal configuration requires fewer API calls, indicating that visual modality enables the agent to localize relevant code regions with fewer iterative navigation and verification steps compared to the text-only baseline.

Taken together, these results suggest that multimodal context integration improves issue resolution not by replacing textual reasoning, but by guiding repository exploration. The structural visualization provides a global view of repository structure, allowing agents to identify relevant regions earlier and avoid redundant navigation steps. As a result, reasoning trajectories become shorter and more focused, reducing token consumption while maintaining repair accuracy across different models. 
The magnitude of token reduction varies across models, reflecting differences in exploration and tool-usage behaviors.

As shown in Tables~\ref{tab:gpt5mini_swerebench} and~\ref{tab:gpt5mini_sweqa}, the multimodal setting preserves or slightly improves effectiveness on both benchmarks while reducing interaction cost. On SWE-Rebench Leaderboard, \approach{} raises Pass@1 from 25.45\% to 26.36\% while cutting input tokens by 34.89\% and cost by 9.6\%. This result indicates that the additional visual information does not compromise the model's repair capability and can even lead to a modest improvement in task performance. On SWE-QA, it improves the Overall Score from 66.8 to 67.2 while reducing API calls by 35.7\% and cost by 26.2\%. The larger efficiency gain on SWE-QA is consistent with the role of structural visualization in localization-heavy tasks: question answering mainly requires identifying the relevant code region rather than completing the full repair pipeline.

\begin{findingbox}
Multimodal context integration reduces token cost by up to 46\% on SWE-bench Verified while maintaining or improving accuracy, and the same effectiveness-efficiency trend transfers to additional GPT-5-mini evaluations on SWE-Rebench and SWE-QA.
\end{findingbox}

\section{RQ3: Effect of Visual Layout}
With the efficiency of multimodal integration verified, we further investigate which visual layout is the most effective for coding agents. We compare three visual rendering of repository, including graph, nested, and tabular. Additionally, we examine the effect of hierarchy depth within the visual representations. 

\subsection{Experimental Design}

We compare three visual rendering strategies, as illustrated in Figure~\ref{fig:layouts}. The three variants render the repository subgraph as images.
\textbf{Graph} renders the subgraph as a directed graph where nodes are connected by edges. Node types are distinguished using different icons (e.g., folder icons for directories, document icons for files), while dependency directions are encoded through arrow orientation, preserving the full topological structure of the dependency graph.
\textbf{Nested} extends the graph layout by grouping nodes belonging to the same directory or file within dashed bounding boxes, making hierarchical containment spatially explicit without requiring the agent to trace \textit{contains} edges.
\textbf{Tabular} removes explicit edges entirely and presents nodes as a flat, color-coded list: the query node in yellow, its parent directory in blue, subdirectories in purple, and contained files grouped in white blocks, encoding relational context through color rather than topology.
In addition to the visual layouts, we consider a textual rendering strategy as a reference. 
\textbf{Text} linearizes the repository structure queried by \approach{} into a sequential textual format (e.g., listing nodes and their relationships as structured text). 
\begin{table}[t]
  \centering
  \caption{Comparison of visualization layout on SWE-bench Verified (GPT-5 mini, 500 instances).}
  \footnotesize
  \setlength{\tabcolsep}{4pt}
  \begin{tabular}{lcccc}
    \toprule
    \textbf{Method} & \textbf{Pass@1} & \textbf{Input Tokens} & \textbf{Output Tokens} & \textbf{Cost (\$)} \\
    \midrule
    Mini-SWE-Agent        & 55.0\% & 193,157 & 8,188 & 0.031 \\
    \addlinespace[2pt]
    + Text        & 53.8\% & 159,558~{\textcolor{darkgreen}{($-$17\%)}} & 8,380~{\textcolor{darkred}{(+2\%)}} & 0.027~{\textcolor{darkgreen}{($-$12\%)}} \\
    + Graph     & 55.4\% & \textbf{144,403}~{\textcolor{darkgreen}{($-$25\%)}} & \textbf{6,958}~{\textcolor{darkgreen}{($-$15\%)}} & \textbf{0.023}~{\textcolor{darkgreen}{($-$26\%)}} \\
    + Nested      & 55.8\% & 154,788~{\textcolor{darkgreen}{($-$20\%)}} & 7,458~{\textcolor{darkgreen}{($-$9\%)}} & 0.025~{\textcolor{darkgreen}{($-$18\%)}} \\
    + Tabular     & \textbf{56.2\%} & 163,311~{\textcolor{darkgreen}{($-$16\%)}} & 7,545~{\textcolor{darkgreen}{($-$8\%)}} & 0.027~{\textcolor{darkgreen}{($-$14\%)}} \\
    \bottomrule
  \end{tabular}
  \label{tab:graph_versions}
\end{table}
\noindent\textbf{Hierarchy Depth.}
When the agent queries the graph tool, it specifies a query node and a traversal depth $k$. The graph tool returns all nodes reachable within $k$ hops along the specified edge type. A larger $k$ exposes a broader dependency neighborhood but also increases the size of the rendered image and the resulting input token count. To study this tradeoff, we fix the traversal depth at $k \in \{1, 2, 3, 4\}$ and compare these settings with \approach{}, where the agent dynamically determines $k$ for each query. All experiments use GPT-5-mini on 500 instances from SWE-bench Verified. The Graph layout with agent-decided depth corresponds to the default \approach{} configuration evaluated in RQ2. 

\subsection{Results and Analysis}
As shown in Table~\ref{tab:graph_versions}, all three visual layouts improve over the text-only baseline. The text representation reduces input tokens by 17\% but marginally hurts accuracy ($-$1.2), indicating that structured text alone cannot fully substitute for visual structural context. Among visual layouts, graph layout achieves the best token efficiency ($-$25\% input tokens, $-$26\% cost) at a modest accuracy gain (+0.4), while tabular layout yields the highest Pass@1 (56.2\%, +1.2) at a lower efficiency gain ($-$16\% cost). Nested layout sits between the two (Pass@1 55.8\%, $-$18\% cost). This tradeoff suggests that graph layout encodes structural relationships more compactly for token-efficient navigation, whereas tabular layout with semantic color-coding provides richer local context that aids precise localization.

As shown in Table~\ref{tab:rq3_depth}, fixed hop depths all reduce cost relative to the baseline. Depth~1 slightly hurts accuracy ($-$0.6) as the shallow neighborhood may omit key dependencies; deeper depths progressively improve accuracy, with Depth~4 achieving the highest Pass@1 (57.2\%, +2.2). However, fixed depths incur more input tokens as depth increases. \approach{} with agent-decided depth achieves a competitive Pass@1 gain (+0.4) with the lowest input tokens ($-$25\%) and cost ($-$26\%) across all depth configurations, suggesting the agent adaptively selects shallow depths when a narrow context suffices and deeper traversals only when required.

\begin{findingbox}
All visual layouts outperform the text-only baseline, with graph layout achieving the best token efficiency ($-$26\% cost). Agent-decided hop depth achieves a competitive Pass@1 gain (+0.4) while delivering the largest reductions in input tokens ($-$25\%) and cost ($-$26\%) across all depth configurations.
\end{findingbox}

\FloatBarrier
\section{RQ4: Effectiveness of Visualization in Different Stages}
With multimodal integration shown to improve both effectiveness and efficiency, we further investigate \textit{when} visualization is most beneficial within the issue resolution pipeline. The contribution of structural visual context may vary across different stages of problem solving, as early stages of issue resolution primarily involve bug localization, whereas later stages focus on code modification and validation.

\begin{table}[t]
  \centering
  \caption{Effect of hierarchy depth on SWE-bench Verified (GPT-5-mini, 500 instances). \approach uses an adaptive depth determined by agents.}
  \footnotesize
  \setlength{\tabcolsep}{3pt}
  \begin{tabular}{lcccc}
  \toprule
  \textbf{Graph Depth} & \textbf{Pass@1} & \textbf{Input Tokens} & \textbf{Output Tokens} & \textbf{Cost (\$)} \\
  \midrule
  Mini-SWE-Agent     & 55.0\% & 193,157 & 8,188 & 0.031 \\
  \cmidrule(lr){1-5}
  Depth = 1          & 54.4\% \textcolor{darkred}{(-0.6)} & 155,971 \textcolor{darkgreen}{(-19\%)} & 9,577 \textcolor{darkred}{(+17\%)} & 0.028 \textcolor{darkgreen}{(-10\%)} \\
  Depth = 2          & 55.8\% \textcolor{darkgreen}{(+0.8)} & 158,350 \textcolor{darkgreen}{(-18\%)} & 7,782 \textcolor{darkgreen}{(-5\%)} & 0.026 \textcolor{darkgreen}{(-15\%)} \\
  Depth = 3          & 55.4\% \textcolor{darkgreen}{(+0.4)} & 156,019 \textcolor{darkgreen}{(-19\%)} & 7,624 \textcolor{darkgreen}{(-7\%)} & 0.026 \textcolor{darkgreen}{(-16\%)} \\
  Depth = 4          & \textbf{57.2\%} \textcolor{darkgreen}{(+2.2)} & 161,441 \textcolor{darkgreen}{(-16\%)} & 7,466 \textcolor{darkgreen}{(-9\%)} & 0.026 \textcolor{darkgreen}{(-16\%)} \\
  \approach          & 55.4\% \textcolor{darkgreen}{(+0.4)} & \textbf{144,403} \textcolor{darkgreen}{(-25\%)} & \textbf{6,958} \textcolor{darkgreen}{(-15\%)} & \textbf{0.023} \textcolor{darkgreen}{(-26\%)} \\
  \bottomrule
  \end{tabular}
  \label{tab:rq3_depth}
\end{table}

\subsection{Experimental Design}
Issue resolution proceeds through three distinct stages: bug localization, patch repair, and patch validation~\cite{yang2024sweagenta, xia2024agentless}.To isolate the effect of invocation stage, we construct three variants that each confine visualization-tool invocation to a single stage. Specifically, we enforce stage-specific invocation by enabling or disabling the graph tool through the system prompt and tool definitions associated with each phase. The localization variant corresponds to the default \approach{} setting; the other two are ablations that shift invocation to later stages:

\noindent\textbf{Localization Variant.}
To isolate the effect of structural grounding during early issue analysis, the agent is equipped with the \approach{} graph tool only in the localization phase, where it explores repository structure and identifies relevant code entities. All subsequent stages rely solely on standard Bash commands.

\noindent\textbf{Repair Variant.}
To evaluate the role of structural information during code modification, the agent uses standard Bash commands for navigation and editing, but invokes the \approach{} graph tool before applying changes. The tool is used to inspect upstream and downstream dependencies of the target entity, helping avoid unintended side effects during patch construction.

\noindent\textbf{Patch Validation Variant.}
To examine whether structural grounding primarily benefits post-modification verification, the agent performs localization and repair using standard Bash commands only. After generating a patch, the agent invokes the \approach{} graph tool to inspect the dependency neighborhood of the modified entity before generating and executing validation tests.

\begin{table}[H]
  \centering
  \caption{Effect of visualization in different stages on SWE-bench Verified (GPT-5-mini, 500 instances). The percentages in parentheses indicate relative change to Mini-SWE-Agent.}
  \label{tab:rq4_timing}
  \footnotesize
  \setlength{\tabcolsep}{3pt}
  \begin{tabular}{lc|ccc}
  \toprule
  \textbf{Stage} & \textbf{Pass@1} & \textbf{Input Tokens} & \textbf{Output Tokens} & \textbf{Cost (\$)} \\
  \midrule
  Mini-SWE-Agent   & 55.0\% & 193,157 & 8,188 & 0.031 \\
  \cmidrule(lr){1-5}
  Localization     & \textbf{55.4\%}~{\textcolor{darkgreen}{(+0.4)}} & \textbf{144,403}~{\textcolor{darkgreen}{($-$25\%)}} & \textbf{6,958}~{\textcolor{darkgreen}{($-$15\%)}} & \textbf{0.023}~{\textcolor{darkgreen}{($-$26\%)}} \\
  Repair           & 50.0\%~{\textcolor{darkred}{($-$5.0)}} & 174,544~{\textcolor{darkgreen}{($-$10\%)}} & 8,879~{\textcolor{darkred}{(+8\%)}} & 0.029~{\textcolor{darkgreen}{($-$5\%)}} \\
  Patch Validation & 51.6\%~{\textcolor{darkred}{($-$3.4)}} & 178,922~{\textcolor{darkgreen}{($-$7\%)}} & 8,758~{\textcolor{darkred}{(+7\%)}} & 0.030~{\textcolor{darkgreen}{($-$4\%)}} \\
  \bottomrule
  \end{tabular}
\end{table}

\begin{figure*}[t]
  \centering
  \includegraphics[width=\textwidth,trim=0 0 0 0,clip]{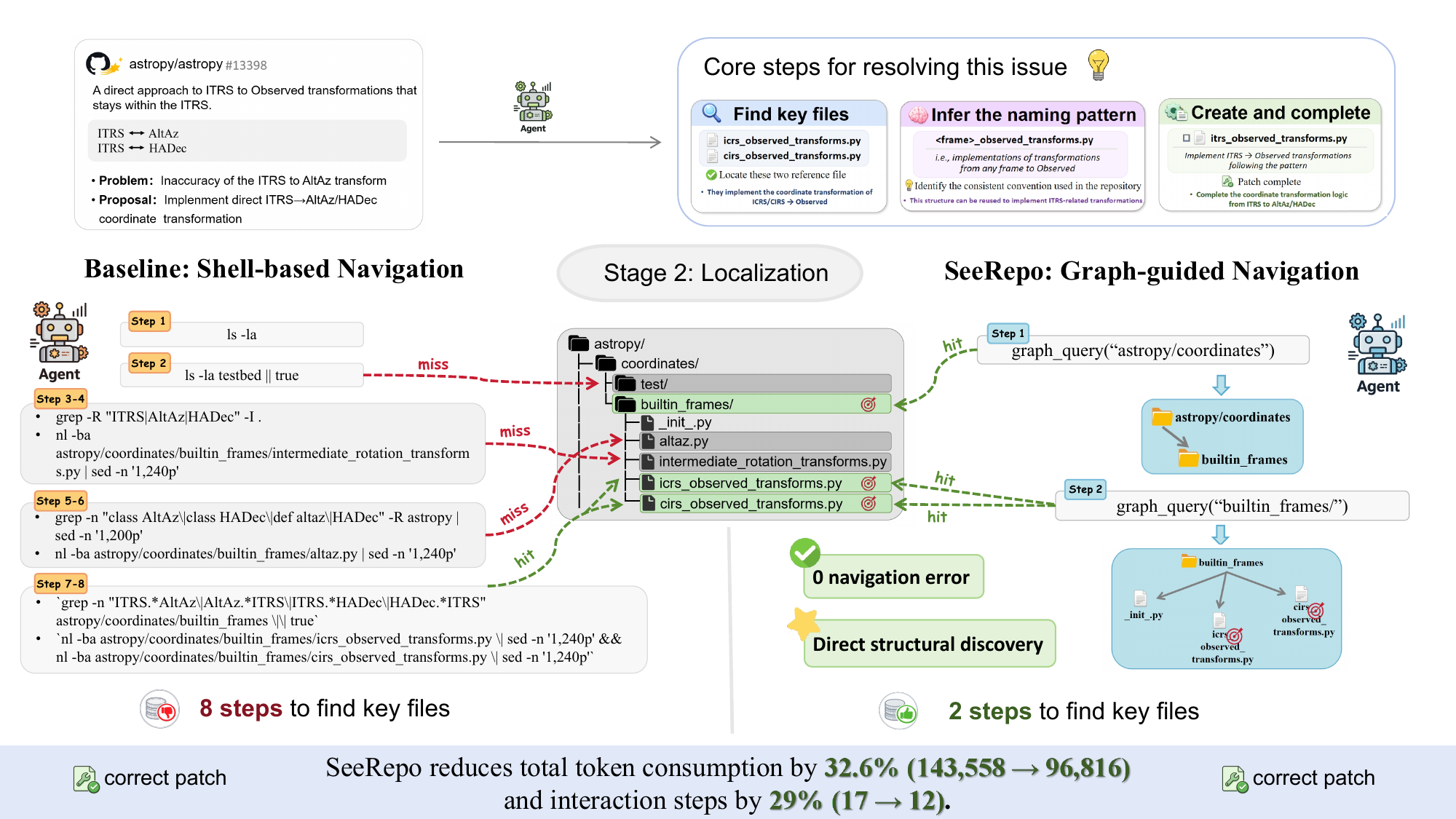}
  \caption{Case study on SWE-bench Verified instance \texttt{astropy\_\_astropy-13398}.}
  \label{fig:case}
\end{figure*}

\subsection{Results and Analysis}
As shown in Table~\ref{tab:rq4_timing}, the effectiveness of visualization varies substantially depending on the stage at which it is invoked. Enabling visualization during the repair stage yields a Pass@1 of 50.0\% ($-$5.0) while providing only a marginal 5\% reduction in cost. After localization has already identified candidate files, invoking the graph tool during repair exposes the agent to a broader set of upstream and downstream dependencies, many of which are not directly relevant to the target fix. This additional structural context appears to introduce distraction, interfering with the precise textual reasoning required for accurate code editing.

Deferring visualization to the patch validation stage partially recovers performance (51.6\%, $-$3.4) but still underperforms the baseline. At this point, structural information is introduced only after a patch has been produced; inspecting dependency neighborhoods may surface seemingly related components and encourage unnecessary follow-up modifications, potentially expanding patch scope and increasing the likelihood of regressions.

In contrast, introducing \approach{} to the localization stage produces the best overall results, achieving a Pass@1 of 55.4\% (+0.4) together with a 25\% reduction in input tokens and a 26\% decrease in cost. During early bug localization, visual modality helps the agent narrow the candidate search space and identify relevant code entities prior to repair, reducing redundant exploration while preserving accurate trajectory reasoning.

\begin{findingbox}
Visualization is most effective at the fault localization stage: \approach{} achieves Pass@1 of 55.4\% (+0.4) with 26\% cost reduction.
\end{findingbox}

\section{Case Study}

To qualitatively illustrate how \approach reduces exploratory overhead, we showcase the instance \texttt{astropy\_\_astropy-13398} from SWE-bench Verified, which requires implementing direct ITRS$\leftrightarrow$\allowbreak AltAz and ITRS$\leftrightarrow$HADec coordinate transformations in the \texttt{astropy} library by creating \texttt{itrs\_observed\_transforms.py} and registering it in \texttt{builtin\_frames/\allowbreak\_\_init\_\_.py}. Figure~\ref{fig:case} contrasts the localization trajectories of the baseline agent and \approach on this instance.

Without structural context, the baseline agent resorts to an iterative grep-then-read strategy. After initial repository inspection via \texttt{ls~-la} (Steps~1--2), it launches a broad keyword search (\texttt{grep~-R~"ITRS|AltAz|HADec"}) across the entire repository, which returns voluminous but low-relevance output (Step~3, \emph{miss}). The agent then reads \texttt{intermediate\_rotation\_transforms.py} to infer the frame structure (Step~4, \emph{miss}), and further issues targeted class-level searches (\texttt{grep~-n~"class~AltAz\textbackslash|class~HADec..."}) followed by reading \texttt{altaz.py} (Steps~5--6, \emph{miss}). Only at Steps~7--8, after additional pattern-matching queries, does the agent finally locate \texttt{icrs\_observed\_transforms.py} and \texttt{cirs\_observed\allowbreak\_transforms.py}---the template files that reveal the naming convention and registration mechanism needed for the fix. In total, eight interaction steps are consumed before the key files are identified, with three rounds of \emph{miss} preceding the eventual \emph{hit}.

In contrast, the multimodal agent equipped with \approach enables the agent to reach the same structural understanding in two steps. The agent first issues \texttt{graph\_query\allowbreak
("astropy/coordinates")} (\allowbreak Step~1), which returns the \texttt{contains}-edge subgraph rooted at this directory and immediately surfaces \texttt{builtin\_\allowbreak frames/} as a relevant subdirectory (\emph{hit}). A follow-up query \texttt{graph\_query("builtin\_\allowbreak frames/")} (Step~2) retrieves the complete file roster of that directory (\emph{hit}), directly exposing the naming pattern \texttt{*\_observed\_\allowbreak transforms.py} instantiated by \texttt{cirs\_\allowbreak
observed\_transforms.py} and \texttt{icrs\_\allowbreak observed\allowbreak \_transforms.py}. This structured response provides both an implementation template and implicit guidance on the registration mechanism, allowing the agent to proceed to creating \texttt{itrs\_observed\_\allowbreak transforms.py} with zero navigation errors.

While both agents ultimately produce correct patches, \approach reduces total token consumption by 32.6\% (143{,}558 $\rightarrow$ 96{,}816) and interaction steps by 29\% (17 $\rightarrow$ 12). The efficiency gap originates in the localization stage: the baseline's repeated grep outputs and file reads accumulate approximately 25K tokens of low-information-density context before productive modification begins, whereas two graph queries provide equivalent---and more structured---information at a fraction of the token cost.

This case illustrates the core benefit of \approach: by replacing trial-and-error shell exploration with topology-aware queries, agents can devote more of their limited context window to reasoning and code modification rather than navigation.

\section{Discussion}

\subsection{Threats to Validity}

\textbf{External Validity.}
Our evaluation is conducted on SWE-bench Verified, which consists exclusively of Python repositories. Although this benchmark provides realistic issue-resolution scenarios with reliable test-based validation, the observed benefits of structural visualization may depend on characteristics specific to Python projects, such as module organization patterns and dependency structures. Consequently, it remains unclear whether the same efficiency and reasoning improvements would generalize to repositories written in other programming languages (e.g., Java or TypeScript) that exhibit different architectural conventions or build systems.

\textbf{Construct Validity.}
We measure efficiency primarily through token consumption and reasoning trajectory length, using them as proxies for exploration efficiency. While these metrics capture computational cost and navigation behavior, they may not fully reflect qualitative aspects of reasoning, such as interpretability or developer-aligned debugging strategies. Future studies incorporating human evaluation or finer-grained behavioral analyses could provide a more comprehensive assessment of agent reasoning quality.

\textbf{Internal Validity.}
To isolate the effect of structural visualization, all experimental settings are kept identical to the text-only baseline except for the addition of structural context. Nevertheless, interactions between visualization and specific model architectures may still influence outcomes. For example, models with different planning or tool-usage tendencies may exploit structural context to varying degrees, potentially affecting token reduction ratios across models.

\subsection{Future Work}
Several promising directions remain for future exploration.
First, the current structural visualization relies on static Graphviz layouts, which may become visually dense when applied to large-scale repositories with complex dependency structures. Developing adaptive visualization strategies that dynamically emphasize query-relevant subgraphs or progressively reveal structural information could substantially improve interpretability and scalability.

Second, while our current framework allows agents to decide the depth of structural exploration, more principled mechanisms for controlling visualization scope remain unexplored. Learning-based invocation policies, such as reinforcement learning or uncertainty-aware triggering mechanisms, could enable agents to request structural context only when it is expected to provide measurable reasoning benefits. 

Finally, extending structural grounding beyond static analysis is an important avenue for future work. Incorporating dynamic signals, such as execution traces or runtime dependencies, may enable richer representations of repository behavior and further enhance agent reasoning in complex software environments. Such hybrid representations could also help agents distinguish between frequently executed code paths and rarely triggered branches, enabling more targeted localization and repair.

\section{Related Work}
\textbf{Software Engineering Agents.} Recent years have seen rapid progress on LLM-based agents~\cite{chang2026test,peng2025swe,wang2026swe,lin2024llms,wang2026context,shi2025longcodezip,shi2024code,shi2025between,zhang2026swe,lin2026knowfixqadrivenrepository, hu2025flowmaltransunsupervisedbinarycode, hu2026zeroshotvulnerabilitydetectionlowresource} for repository-level issue resolution. On the scaffold side, SWE-agent~\cite{yang2024sweagenta} demonstrates that an agent-computer interface (ACI) that supports repository navigation, editing, and execution can substantially improve task performance. AutoCodeRover \cite{zhang2024autocoderover} further incorporates software engineering oriented context retrieval, leveraging an AST-based program representation (e.g., class/method structure) and iterative search to ground patch generation for GitHub issues. Recent work also improves repository navigation and localization using structured signals and scheduling: LocAgent~\cite{chen2025locagent} uses graph-guided multi-hop traversal to localize relevant entities, RepoMem~\cite{wang2025repomem} augments localization with repository memory mined from history, and OrcaLoca~\cite{yu2025orcaloca} improves localization with scheduling and distance-aware context pruning. Building on such scaffolds, experience-driven approaches aim to reuse past repair knowledge rather than treating each issue in isolation: SWE-Exp~\cite{chen2025swe} constructs an experience bank from historical trajectories to guide planning and patching, while ExpeRepair~\cite{mu2025experepair} introduces a dual-memory design (episodic demonstrations and semantic reflections) to dynamically compose prompts for repository-level repair. Orthogonally, test-time scaling and search-based methods increase inference-time compute to explore and refine candidate solutions: SWE-Debate~\cite{li2025swe} uses competitive multi-agent debate (and integrates search during patching) to improve fault localization and fix planning, and SWE-Search~\cite{antoniades2024swesearch} augments agents with Monte Carlo Tree Search and iterative refinement to enable backtracking and deeper exploration. Relatedly, SAGE~\cite{hayashi2025self} improves agent behavior by self-abstracting grounded experience into compact plans for subsequent re-execution, providing a complementary path to boost long-horizon performance at test time. In contrast to improving agents via new scaffolds, memories, or inference-time search, our work is the first to study multimodal representations of code repositories as a design dimension for SWE agents. We find that representing repository structure as visual graph images consistently reduces token cost while maintaining resolution accuracy, and that the benefit is most pronounced when visualization is invoked at the localization stage.

\textbf{Multimodal Coding Agents.} As more real-world software issues are reported with visual evidence (e.g., screenshots), recent benchmarks have begun to evaluate agents in software engineering with visual element information. SWE-bench Multimodal (SWE-bench M)~\cite{yang2025swebench} extends SWE-bench to visual, user-facing JavaScript repositories by providing issue resolution tasks that include images in problem statements or tests, enabling evaluation in visual software domains. Building on this benchmark, recent work has started to incorporate multimodal signals into coding agents and automated repair. GUIRepair~\cite{huang2025seeingisfixing} studies visual software issue fixing by enabling cross-modal reasoning between GUI screenshots and code, and by using rendered visual feedback to support patch validation. OpenHands-Versa~\cite{soni2025codingagents} augments coding agents with multimodal browsing as a generalist capability and evaluates on benchmarks including SWE-bench Multimodal~\cite{yang2025swebench}. SVRepair~\cite{tang2026svrepair} proposes structured visual reasoning for automated program repair by transforming heterogeneous visual artifacts into semantic scene graphs to guide localization and patch synthesis. 
Beyond incorporating external visual artifacts, recent work has also explored representing code itself through visual modalities. CodeOCR~\cite{shi2026codeocr} renders source code as images to enable multimodal models to process programs with improved token efficiency while maintaining performance on code understanding tasks. This line of work suggests that visual representations can serve as an alternative interface for presenting software information to foundation models.
In contrast, \approach{} adopts a multimodal design that visualizes repository structure rather than code content. Instead of encoding individual files as images, \approach{} constructs structural visualizations derived from static analysis, exposing dependency relationships and global repository organization that are often implicit in text-only interaction. In this formulation, repository structure is presented through a visual modality, while fine-grained code details remain in their original textual form for precise semantic reasoning. Our work therefore examines how structural visual context affects issue resolution in realistic software engineering (SWE) tasks, providing a complementary perspective on multimodal representations beyond code-centric visual encoding.

\balance
\section{Conclusion}
This paper presents the first systematic empirical study of visual repository representations for MLLM-based software engineering agents on repository-level issue resolution. We introduce \approach{}, a multimodal framework that presents different types of repository information through appropriate modalities: structural and dependency relationships are rendered as visual graph images, while code content is retained as text. This design leverages the complementary strengths of MLLMs---visual perception for structural orientation and symbolic reasoning for precise code understanding---enabling agents to navigate large repositories more effectively.

Our experiments on SWE-bench Verified with four models yield four key findings. First, vision-only modality is insufficient: replacing text access with graph images degrades accuracy by up to 34.1 points while paradoxically inflating token cost. Second, multimodal integration---adding \approach{} alongside standard text tools---reduces cost by up to 46\% across all models while maintaining or improving resolution accuracy. Third, among visual layout designs, graph layout provides the best token efficiency and agent-decided hop depth achieves the best cost reduction while maintaining competitive accuracy. Fourth, visualization is most effective at the localization stage; invoking it during repair or validation degrades performance due to noise introduction and scope expansion.

These findings collectively suggest that visual repository representations are a practical and cost-effective complement to text-based interaction for coding agents, provided they are designed and invoked appropriately. More broadly, our results indicate that the modality through which structural information is presented---not merely its content---meaningfully shapes agent behavior and cost. We hope this work motivates further exploration of multimodal representations in the development of future coding agents.

\begin{acks}
This research is funded by the National Key Research and Development Program of China (Grant No. 2023YFB4503802) and the Natural Science Foundation of Shanghai (Grant No. 25ZR1401175).
\end{acks}

\section*{Data Availability Statement}
All data and artifacts are publicly available at \url{https://doi.org/10.5281/zenodo.19229174}.

\bibliographystyle{ACM-Reference-Format}
\bibliography{ref}

\end{document}